\documentclass[12pt]{article}
 \usepackage{latexsym,epsfig,color,times}
\usepackage{colordvi}
\usepackage{amsmath}
\usepackage{latexsym,graphicx,color,times}
\usepackage[portrait,margin=1in]{geometry}
\def\Rdd{\Red}
\def\Rdd{\Black}

\def\Grn{\Cyan}
\def\Grn{\Black}

\def\pd{\partial}

\def\etal{\it{ et al.} \em }
\def\Alf{Alfv\'en }
\def\rfig#1{Fig.\ref{fig:#1}}
\def\rsec#1{Sec.\ref{sec:#1}}

\newcommand{\be}{\begin{equation}}
\newcommand{\ee}{\end{equation}}
\def\pd{\partial}

\def\pd{\partial}

\def\Alf{Alfv\'en }

\def\req#1{(\ref{eq:#1})}

\def\macname{hank_strauss}

\def\figdird3d{/Users/{\macname}/Documents/progs/m3d/d3d-lock}
\def\figditer{/Users/{\macname}/Documents/progs/m3d/itlock}
\def\figdirw{/Users/{\macname}/Documents/papers/disruption/rwtm/gnuplot}

\def\figdirp5{/Users/{\macname}/Documents/papers/disruption/rwtm/deltap/paper5}

\def\figdirden{/Users/{\macname}/Documents/papers/disruption/rwtm/mst-density}
 
\begin{document}

\begin{center}
\vspace{.5cm}
%\title{\raggedright
%  \textbf{\fontsize{12}{15}\selectfont
%   {\textbf{\fontsize{15}
%{\Large\bf Resistive Wall Tearing Modes in Disruptions } %, \\
%{\Large\bf Current Contraction and Feedback in Resistive Wall Tearing Mode Disruptions } %, \\
{\Large\bf Resistive Wall Tearing Mode Disruptions } %, \\
%{\Large\bf Resistive Wall Tearing Mode Disruptions and Current Contraction } %, \\
\end{center}

\begin{center}
  H. R. Strauss $^1$, B. E. Chapman $^2$, B. C. Lyons $^3$ \\ 
  $^1$ HRS Fusion, West Orange,  USA \\
\Rdd{$^2$ Dept. of Physics, University of Wisconsin, Madison WI 53706 USA  } \\
\Rdd{$^2$ General Atomics, San Diego, CA 92121 USA}
\end{center}

  Email: hank@hrsfusion.com

\bigskip

\centerline{{\large \bf Abstract}}{\it
This paper deals with resistive wall tearing mode (RWTM) disruptions.
RWTMs are closely related to resistive wall modes (\Rdd{RWMs}). 
The nonlinear
behavior of these modes is strongly dependent on the resistive wall outside the
plasma. 
A conducting wall is  highly  mitigating
for RWTM disruptions. 
The consequence for ITER, \Rdd{which has a highly conducting wall, } is that the thermal quench (TQ) time could  be
much longer than previously conjectured.  %as indicated in  \rfig{itlock}.
Active feedback stabilization is another  possible way to mitigate or  prevent RWTM disruptions.
Simulations of  disruptions are  reviewed for DIII-D and MST. 
MST has a longer resistive wall time than
ITER, and disruptions are not observed experimentally \Rdd{when MST is operated as a standard tokamak}. Simulations indicate that
the RWTM disruption time scale is longer than the experimental shot time. 
 In general, edge cooling by tearing mode \Rdd{island}  overlap or by impurity
radiation causes contraction of the  current 
profile, which  destabilizes RWTMs.
The equilibria studied here have a $q = 2$ rational surface close to
the edge of the plasma, and low current density between the $q = 2$ surface and
the wall.  A sequence of low edge current model equilibria has 
major disruptions only for a resistive, not ideal, wall, and
edge $q \le 3.$
This is  consistent with typical regimes of tokamak
disruption avoidance, suggesting that  typical tokamak disruptions
could be  RWTMs. 
}

  \begin{section}{Introduction} \label{sec:intro}
Disruptions are loss of plasma confinement in  tokamaks,
which  could  damage large tokamaks like ITER.
The thermal flux in an ITER disruption
would be intolerable if it occurred on the timescale typical of \Rdd{most}  present tokamaks.
Until recently, the instability which caused locked mode disruptions was not known.
Recent work identified the  thermal quench  in JET  locked mode disruptions  with
a resistive wall tearing mode (RWTM)  \cite{jet21}.
The RWTM was also predicted in ITER  \cite{iter21}, where it 
 produces a slow self mitigated thermal quench.
A
similar instability was found  in a  DIII-D locked mode shot %[3,4].
% 154576 
\cite{d3d22,sweeney}.
The MST experiment  \cite{mst23,hurst} 
\Rdd{when operated as a standard tokamak,}
does not have disruptions. 
Recent theory and  simulations  showed this is because the timescale of  RWTMs, which could cause
a thermal quench,  is longer than the experimental pulse time.

The RWTM instability is studied with simulations, theory, and comparison to experimental data.
%Linear theory is extended to include resistive wall modes with a rational surface in the plasma.
Linear simulations show the mode is stable for an ideal wall, and
unstable with a resistive wall.
Nonlinear simulations show that the
mode grows to large amplitude, causing a thermal quench.
The mode onset occurs when  the rational
surface is sufficiently close to  the plasma edge \cite{sweeney2017}, and the edge current density 
is sufficiently small.

%It is possible to prevent  disruptions caused by RWTMs with
%active feedback, as shown by simulations in progress.

   RWTM disruptions can be passively slowed by a highly conducting wall,
or actively slowed by feedback. Simulations of feedback stabilization of RWTM disruptions
will be presented.

Simulations of a DIII-D disruption show the dependence of the linear growth rate
on $\gamma \propto S_{wall}^{-2/3},$ for large $S_{wall} = \tau_{wall} / \tau_A,$  
where $\tau_{wall}$ is the resistive wall penetration time, 
$\tau_{wall} = r_w \delta_w / \eta_w,$ $r_w$ is the wall minor radius, $\delta_w$ is the wall thickness, $\eta_w$ is the
wall resistivity, 
and $\tau_A$ is the \Alf time. 
In the simulations, the RWTMs %grow to large amplitude at the linear growth rate, causing
%a complete thermal quench. 
\Grn{cause a complete thermal quench, in a time proportional to the linear growth time.}
There is good agreement between the simulations and experimental data.

Simulations of MST show that RWTMs connect smoothly to RWMs, when the edge $q_a =
m / n$, where $(m,n)$ are poloidal and toroidal mode number. For all $q_a$, the RWTM  growth rate is
proportional to  $S_{wall}^{-1},$  and even satisfies the same \Grn{linear} dispersion relation as a RWM.
This scaling holds when $S^{3/4}S_{wall}^{-5/4}$ is sufficiently small, 
where $S$ is the plasma Lundquist number. This might be satisfied in ITER.

The onset condition for RWTMs requires the $q = 2$ rational surface to be
sufficiently close to the wall. It also requires the  current density near the plasma edge  
to be  small. Experimentally, disruptions are often preceded by edge cooling, which
causes the current profile to contract. This can happen in locked mode disruptions and
density limit disruptions. 
A sequence of model low edge current equilibria is simulated, showing that \Grn{major} disruptions
occur for small enough $q_a \approx 3.$  
Only minor
disruptions occur if the wall is ideally conducting. Major disruptions only occur
when the wall is resistive, indicating that they are RWTMs.

RWTMs can be stabilized passively by a sufficiently conducting wall, which is
sufficiently close to the plasma. It appears that active feedback stabilization
is also possible. This is illustrated in simulations.

\Rdd{
 \rsec{tq}  reviews RWTM disruptions, including observed or predicted 
TQ times for JET, DIII-D, ITER, and
MST based on experimental data, theory, and simulations. It suggests a difference in
cases with
high and low $S_{wall}.$  
\rsec{d3d}  summarizes DIII-D locked mode studies.
\rsec{mst}  reviews MST disruptions, which are not observed due to the short pulse time 
but are predicted by
theory and simulations, and discusses the relation between RWTMs and RWMs. 
\rsec{onset}  describes current contraction as a precursor to  disruptions,
with nonlinear studies of contracted equilibria including the
effect of a resistive or ideal  walls.
\rsec{feedback}  presents  work on feedback stabilization; 
and \rsec{conclusion}  gives conclusions.}

\begin{section}{RWTM disruptions} \label{sec:tq}
In devices with long resistive wall
magnetic perturbation time, the TQ  duration produced by RWTMs  is long.
 \begin{figure}[h]
\vspace{.5cm}
\begin{center}
\hspace{.25cm}
 \includegraphics[height=5.0cm]{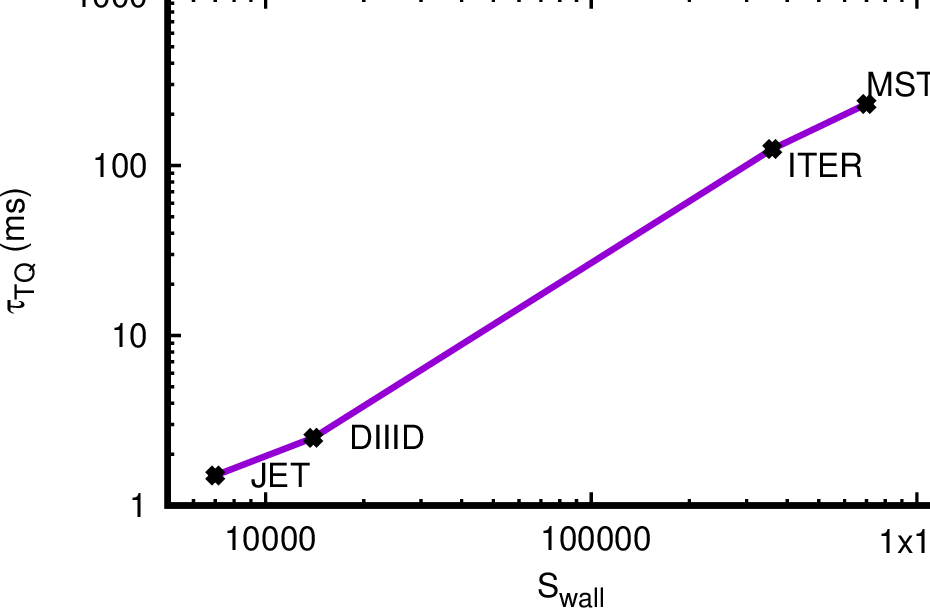}
\end{center}
\caption {\it
  Experimental and simulated thermal quench time 
  $\tau_{TQ}$  in ms, as a function of $S_{wall} = \tau_{wall}/\tau_A,$ 
 indicating a much longer TQ in ITER and MST than in JET and DIII-D.
% (Reproduced from \cite{mst23} with permission.)
}
 \label{fig:itlock}
 \end{figure}

 \end{section}  

%\begin{section}{TQ time in JET, DIII-D, ITER and MST} \label{sec:tq}
 %\rfig{itlock} 
 \rfig{itlock} shows the TQ duration $\tau_{TQ}$ as a function of
$S_{wall}.$ %where $\tau_{TQ}$ is the TQ duration and  $\tau_A$ is the \Alf time. 
For JET and DIII-D, $\tau_{wall} = 5ms.$
The TQ duration was obtained from
experiment and simulations.
In ITER, $\tau_{wall} = 250ms$  and in MST $\tau_{wall} = 800ms.$
For ITER and MST,  the TQ value is based on theory and
simulations. MST disruptions are not observed experimentally within the
the experimental shot time of $50ms,$ which sets a lower bound on $\tau_{TQ}.$
The simulated $\tau_{TQ} \sim 200 ms.$ There appear to be two regimes of
$\tau_{TQ}$, depending on $S_{wall}.$ In
MST, the RWTM  growth time is proportional to $S_{wall},$ like
a RWM. This may also be the case in ITER. 

%\end{section}
%\begin{wrapfigure}{l}{65mm}\centering
\begin{figure}[h]
\begin{center}
 \includegraphics[height=5.2cm]{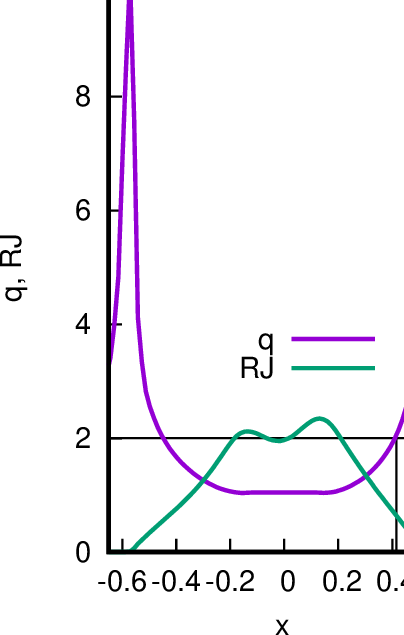}(a)
\includegraphics[width=3.4cm]{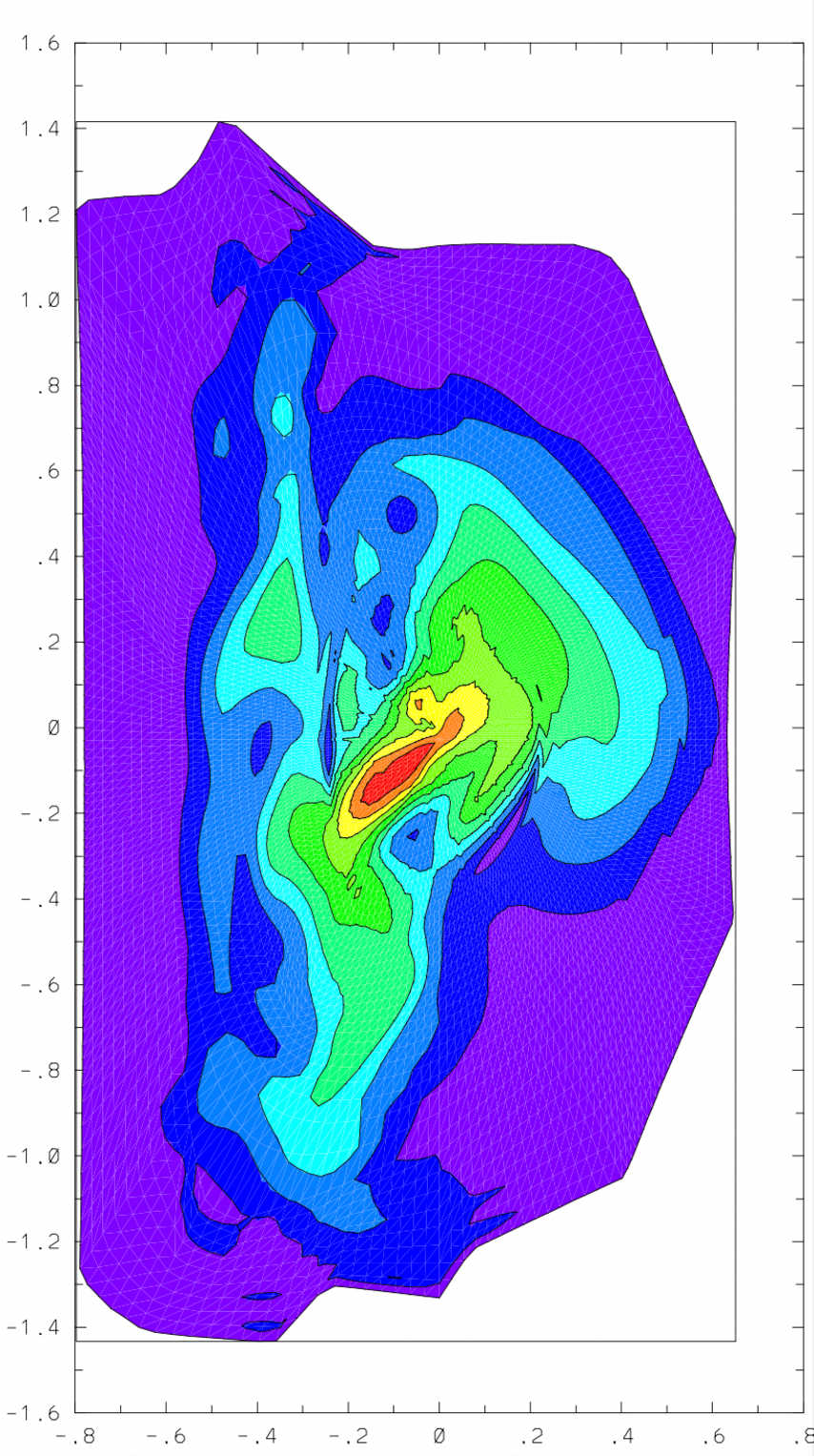}(b)
\includegraphics[height=5.2cm]{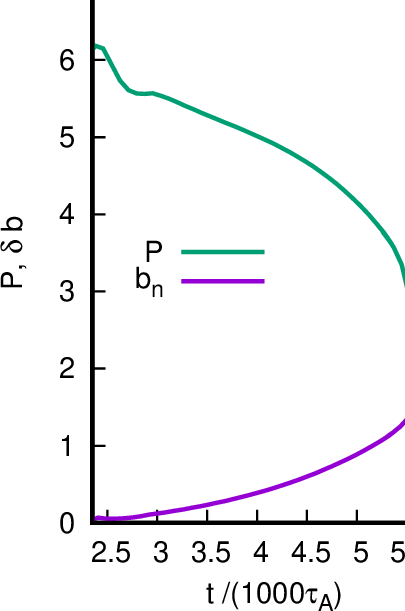}(c)
\end{center}
\caption{\it
(a) initial profiles of $q,$ toroidal current density $RJ_\phi$ in dimensionless units
as a function of $x = R - R_0,$ where $R_0$ is the \Rdd{plasma}  major radius.
(b) contours of pressure $p$ during a nonlinear RWTM %at $t = 5690 \tau_A,$
when volume integrated pressure $P$ is about 20\% of its initial value.
(c) Time history of $P$ and normal magnetic perturbation at the wall  $b_n$.
For the experimental value of $S_{wall},$ the TQ time is in good agreement with experiment.
(\rfig{d3d2}(b) reproduced from \cite{d3d22}.) 
}
 \label{fig:d3d2}
\vspace{.5cm}
\end{figure}

\begin{subsection}{DIII-D locked mode disruption} \label{sec:d3d}
Data from  DIII-D locked mode shot 154576 \cite{sweeney}  
 was compared with simulations.
During the locked mode, moderate amplitude magnetic perturbations are present. Then
an $n = 1 $  magnetic perturbation grows to large amplitude, and the core temperature is
quenched, in time $2.5 ms.$ This is followed by a current quench.

Simulations were carried out using an equilibrium reconstruction $0.3 ms$ before the
start of the mode growth. 
Linear simulations 
 with M3D-C1 \cite{m3dc1}
found instability with a resistive wall, and stability with an
ideally conducting wall, and agreement with linear theory. 
The RWTM growth rate scales asymptotically as
 $\gamma\tau_A \propto S_{wall}^{-2/3}.$
The linear simulations establish that the equilibrium reconstruction is unstable 
with a resistive wall, and stable with an ideal wall. 
Initial profiles are shown in \rfig{d3d2}(a), which indicates that
the $q = 2$ rational surface is close to the plasma edge, and the toroidal current density
is small in the edge. These are typical conditions for RWTM instability. 
The initial profiles of $q$ and
toroidal current $RJ_\phi$ as a function of $x = R - R_0,$ where $R_0$ is the \Rdd{plasma}
major radius.
Nonlinear simulations show that the mode grows to
large amplitude, sufficient to cause a thermal quench.

Nonlinear simulations done with M3D \cite{m3d} show contours of  
the temperature
when the total pressure is $20 \%$ of its initial value, in \rfig{d3d2} (b).
The time history of 
the total pressure $P$ is shown in \rfig{d3d2} (c) demonstrating a TQ.
 The radial  magnetic perturbation $b_n$
at the wall is also shown, showing  that the growth of the RWTM  causes the TQ.
Simulations were done with several values of $S_{wall}.$ It was found that
the TQ time scales as $S_{wall}^{2/3},$ like the linear growth time.
For the experimental value $S_{wall} = 1.2 \times 10^4,$ the TQ time is
$\tau_{TQ} = 2.5 ms, $ in agreement with the data.
The simulations also find that the maximum radial magnetic field at the wall agrees with the
maximum value in the experiment.

\end{subsection}
  \begin{subsection}{MST disruptions} \label{sec:mst}
Disruptions are not observed in
the Madison Symmetric Torus (MST) \cite{mst23,hurst} 
\Rdd{when the device is operated as a standard tokamak}.
\Rdd{(At extremely low density, the discharge can contain a large runaway electron 
component, which will not be considered here.)}
The predicted 
 growth time of RWTMs  is much longer than
the experimental shot duration of $50 ms$,
which gives a lower limit to the possible TQ time.
The predicted linear growth rate is  the same as an RWM, even when the
$q = m/n$  surfaces are inside the plasma. 

The predicted thermal quench time
in MST is much longer than in conventional
tokamaks such as JET and DIII-D, and is longer than a prediction for ITER
based on RWTMs, as shown in \rfig{itlock}. 
%The MST case is for edge $q_a = 2.6$. % described in more detail below.
  Simulations were done with M3D of equilibrium reconstructions with
edge $q_a = 2.6, 2,0, 1.7.$ Initial profiles of $q$ and $RJ_\phi$ are shown in  
\rfig{msttqs}(a). They have the typical feature of $q = 2$ rational surface near the
edge, with low edge current.

%In the low edge safety factor  $q_a \le 2.6$ regime of 
In MST, the RWTM
growth time scales linearly in the
resistive wall penetration time \cite{mst23}. %In this regime RWTMs and RWMs satisfy the
This is characteristic of large $S_{wall}$,
in which the RWTM asymptotically satisfies the resistive wall mode (RWM) dispersion relation.
%\rfig{msttqs} (b) shows the TQ time $\tau_{TQ}$ as a function of $q_a.$
%in units of \Alf time $\tau_A,$ for the case $q_a = 2.6.$  The projection to the experimental 
%$S_{wall} = 7 \times 10^5$ of $\tau_{TQ} = 1.5 \times 10^5 \tau_A,$
%or $\tau_{TQ} \approx  180 ms.$
%same linear dispersion relation.
%ITER could also be in this regime.}
%The largest amplitude  magnetic
%perturbation seen in
%simulations is the RWTM with rational surface
%radius $r_s$ at which $q(r_s) = m/n$ is closest
%to $q_a,$ although RWMs can have a comparable amplitude.

The RWTM linear dispersion relation is \cite{mst23,finn95}
\be c_1^{-1}S^{3/4}S_{wall}^{-5/4} ( \hat{\gamma}^{9/4} + g_s \hat{\gamma}^{5/4})
= \Delta_i \hat{\gamma} + g_s \Delta_n  \label{eq:disp1}  \ee
where $\hat{\gamma}  = \gamma \tau_{wall},$ $S$ is the \Rdd{plasma} Lundquist number,
$c_1 = 0.36 m^2 (q_a/2) (q'r_s/q^2)^{1/2}
 \approx 1.7,$ $m$ is the poloidal mode number, $r_s$ is the rational surface radius, $r_w$ is 
the wall radius,
$g_s = 2m/[1 - (r_s/r_w)^{2m}].$
Resistive wall tearing modes have \Rdd{ideal - wall} tearing parameter  $\Delta_{i} \le 0,$
and \Rdd{no - wall} tearing parameter  $\Delta_{n} >  0.$
%and require finite $S_{wall},$
The RWTM growth rate scalings vary as $\gamma \propto S^{-\alpha},$
with $4/9 \le \alpha \le 1.$ In a JET example \cite{jet21}   $\alpha = 4/9,$
while in the DIII-D example  of \rsec{d3d} \cite{d3d22}   $\alpha = 2/3.$ 
In MST simulations  \cite{mst23}  $\alpha = 1.$ This is because of the smallness of
the left side of \req{disp1} $\propto  S^{3/4}S_{wall}^{-5/4} = \sigma.$
%For small $\sigma,$ $\alpha \approx 1.$ %except for $\Delta_i = \order(\sigma).$
Taking edge $S = 10^6,$ 
in JET, with $S_{wall} = 7\times 10^3$, $\sigma = 0.49.$ In DIII-D, with
$S_{wall} = 1.2 \times 10^4$,  $\sigma = 0.25.$ 
 In MST, with  $S_{wall} = 7\times 10^5,$
$\sigma = 1.6\times 10^{-3},$ and 
in   ITER,  with $S_{wall} = 3.5\times 10^5,$  $\sigma = 3.7 \times 10^{-3}.$
ITER can be in the linear $S_{wall}$ regime like MST.

Using a model equilibrium \cite{finn95},
the linear growth rate is,  neglecting the left side of \req{disp1},
\be {\gamma}\Grn{(m,n)} \tau_{wall} =
 - 2m \frac{n q_0 - (m-1)}{nq_0 - (m-1) - (r_0/r_w)^{2m}}. \label{eq:rwm} \ee
This is also the growth rate of a RWM  \cite{finn95}. % ,liu} 
In this model,  $q_0$ is $q$  on axis, the normalized toroidal current density $j = 2/q_0$ is  constant 
within radius $r_0$,  and is zero for larger radius. 
The crossover from RWTM to RWM occurs smoothly
at $q_a  = m/n.$ For $q_a < m/n,$ the rational surface exits the plasma and  the mode becomes a RWM. 
\begin{figure}[h]
\vspace{.5cm}  \begin{center}
\includegraphics[width=7.00cm] {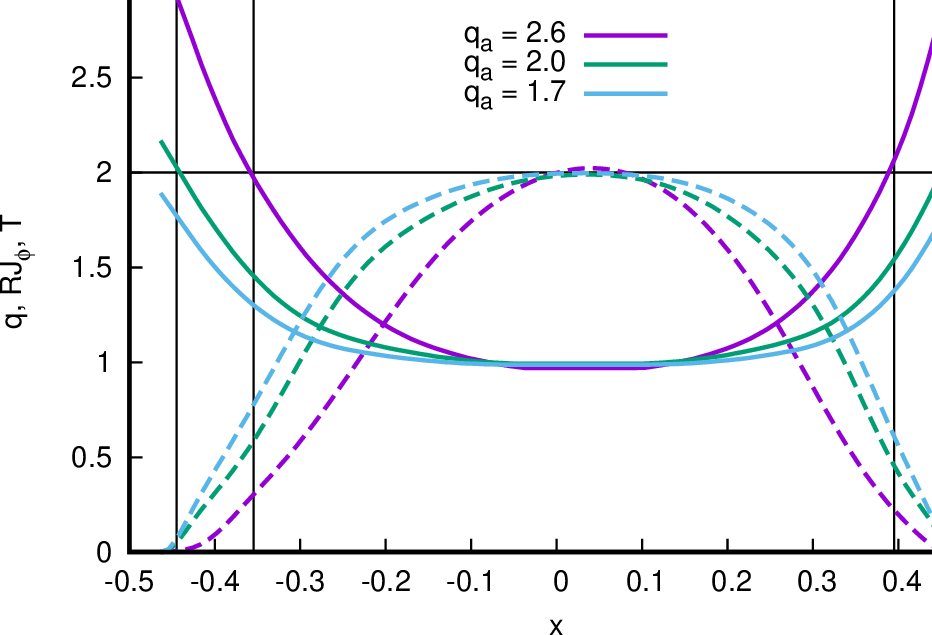}(a)
\includegraphics[width=7.25cm] {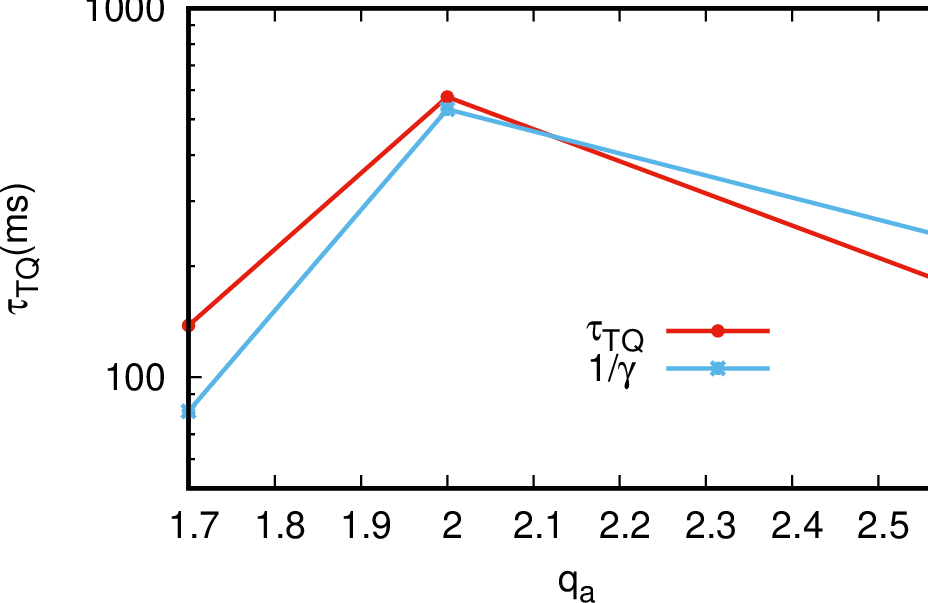}(b)
\end{center}
\caption {\it  
(a) Initial profiles of $q$ and $RJ_\phi$ as a function of $x = R - R_0.$ 
%$\tau_{TQ}/\tau_A$ as a function of $S_{wall},$ for $q_a = 2.6,$ fit approximately by $0.22 S_{wall}.$
(b) $\tau_{TQ}/\tau_A$ as a function of $q_a, $ from the simulations, projected to the experimental
$S_{wall},$ 
and $1 / \gamma$ from \req{rwm}. 
}
\label{fig:msttqs}
\vspace{.5cm}
\end{figure}

\rfig{msttqs}  (b) shows $\tau_{TQ},$ projected to the experimental $S_{wall},$ as a function of
$q_a$ calculated in nonlinear simulations. Also shown is the model growth time 
$1 / \gamma,$ calculated from \req{rwm} with  $q_0 = 1.08$, \Grn{ using the larger growth rate, 
$\gamma(2,1)$ for $q_a \ge 2,$ and $\gamma(3,2)$ for $q_a < 2.$} The agreement is remarkable.
\end{subsection}
\end{section}
 
  \begin{section}{Low edge current and RWTM disruptions} \label{sec:onset}

A common precursor  of disruptions is edge cooling, which causes contraction of the
current profile.
%There can be many sequences of events %such neoclassical tearing modes, 
%leading to a disruption \cite{devries},
%which in JET, usually culminate in a locked mode.
%Locked modes are the main precursor of JET
%disruptions, but they 
%are not the instability causing the
%TQ \cite{gerasimov2020}. %Similarly, edge impurity radiation,
%thermo - resistive tearing modes, could  be precursors, which cause modification of
%It was conjectured \cite{devries16} that at a critical amplitude, tearing modes would
%overlap and cause a disruption,
% although in the JET and DIII-D examples,
%the mode amplitude does not increase before the TQ occurs.
%Prior to and during the locked mode, the plasma can develop
% During disruption precursors such as locked modes, the plasma can develop
During locked modes and other disruption precursors in devices like JET and DIII-D,
low   current and temperature can develop \cite{schuller}
in the plasma edge.
This can be caused by tearing mode island overlap, which has been described \cite{sweeney}  as
a $T_{e,q2}$ collapse, meaning a  minor  disruption causing  a drop of the temperature 
at the $q = 2$ rational surface.
Prior \Rdd{to}  a major disruption, there can be several minor $T_{e,q2}$ disruptions. 
The temperature drop causes the resistivity to increase, and the edge  current is
suppressed. 

Another effect on edge temperature is  impurity radiation 
in the edge or in the core
\cite{pucella}. These effects shape the current profile to be more contracted away from 
the edge, and to become more flattened in the core.
When the edge impurity radiation is large, it 
may be possible to trigger 
\Grn{multifaceted asymmetric radiation from the edge} 
(MARFE) and density limit disruptions   \cite{ferreira2020,lipschultz1984,ricci}.

During  precursors to a disruption,
the plasma edge region cools, causing the current to contract.
A model sequence of low edge current equilibria 
 \cite{model}  
are shown in \rfig{qa3} (a). They were derived from the MST equilibrium with $q_a = 2.6$ in
\rfig{msttqs}(a) 
in which the plasma radius $r_a$  was reduced  so that $r_w / r_a = 1.25,$
as is approximately the case in DIII-D.
Linear combinations of the initial current density  
and the square of the initial current density gave initial states in which $q_0 = 1$ on axis
and $q_a$ has a range of values 
 $ 2 \ge q_a \ge 3.4.$
Current and $q$ profiles as a function of $x = R - R_0$ are shown in \rfig{qa3}(a). 
They have the typical feature of $q = 2$ near the edge, with small edge current density.
 There is a striking
difference in the results, depending on whether the wall is  ideal or resistive.
With an ideal wall,  the perturbations saturate at moderate amplitude, causing a minor
disruption 
without a thermal quench. 
\rfig{qa3} shows time histories of total pressure $P$ cases with $q_a = 2, 2.3, 3.0, 3.4,$ as labeled in the plot. 
With a resistive  wall, indicated with solid curves, there are large perturbations of $P$, which are
major disruptions. For $q_a = 3.4,$ there is a minor disruption.
With an ideal wall, indicated with dotted curves, there are only small perturbations of $P$, 
which are minor disruptions. 
\begin{figure}[h]
\vspace{.5cm}
\begin{center}
\includegraphics[width=7.2cm] {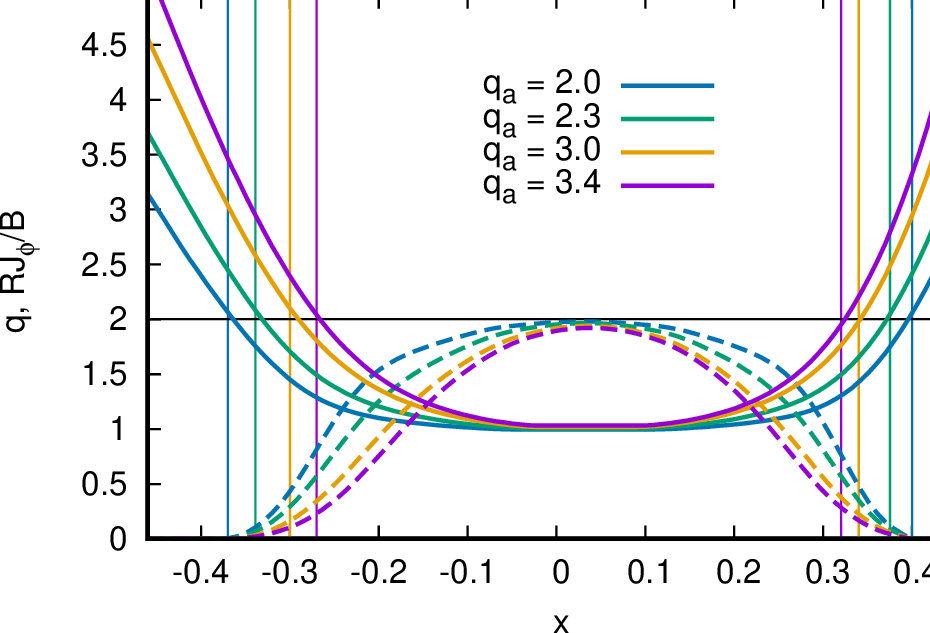}(a)
\includegraphics[width=7.2cm] {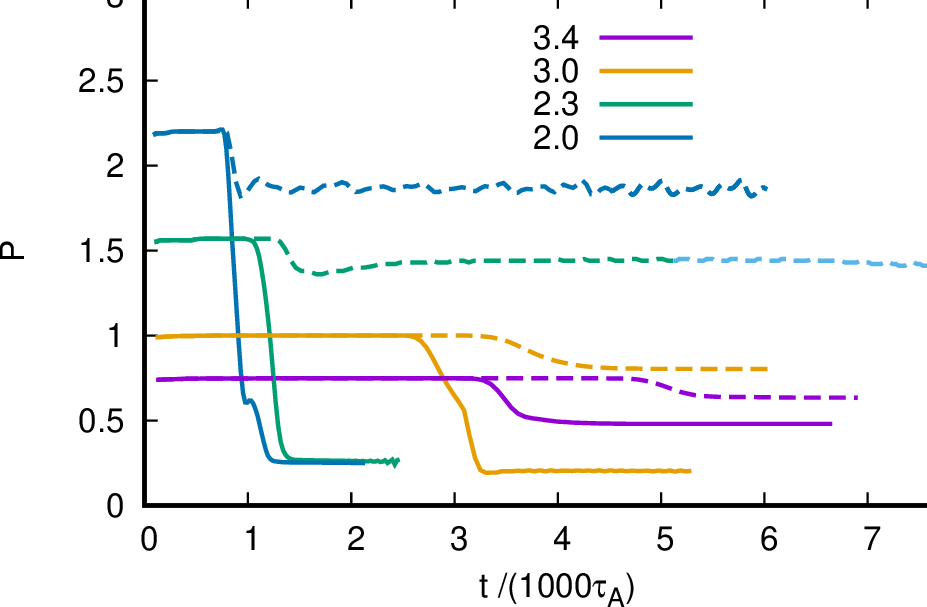}(b)
\end{center}
% \vspace{-.5cm}
\caption {\it
(a) profiles of modified MST equilibria \cite{model}. % arXiv:2308.07694
(b) %TQ time as a function of $q_a,$ calculated from Fig. 8 (b).  %There is an abrupt change in TQ time
Time sequences of total pressure $P$ for initial equilibria with different 
$q_a.$   Solid lines have a resistive wall, while dashed curves have an ideal wall.
There are no major disruptions with an ideal wall, indicating
that major disruptions are RWTMs. For  $q_a = 3.4,$ there are no major disruptions 
even for a resistive wall, indicating a disruption onset boundary for
$3 < q_a \le 3.4.$ 
 }
\label{fig:qa3}
\end{figure}
\vspace{.5 cm}
\begin{figure}[h]
\vspace{.5cm}
\begin{center}
\includegraphics[width=4.3cm]{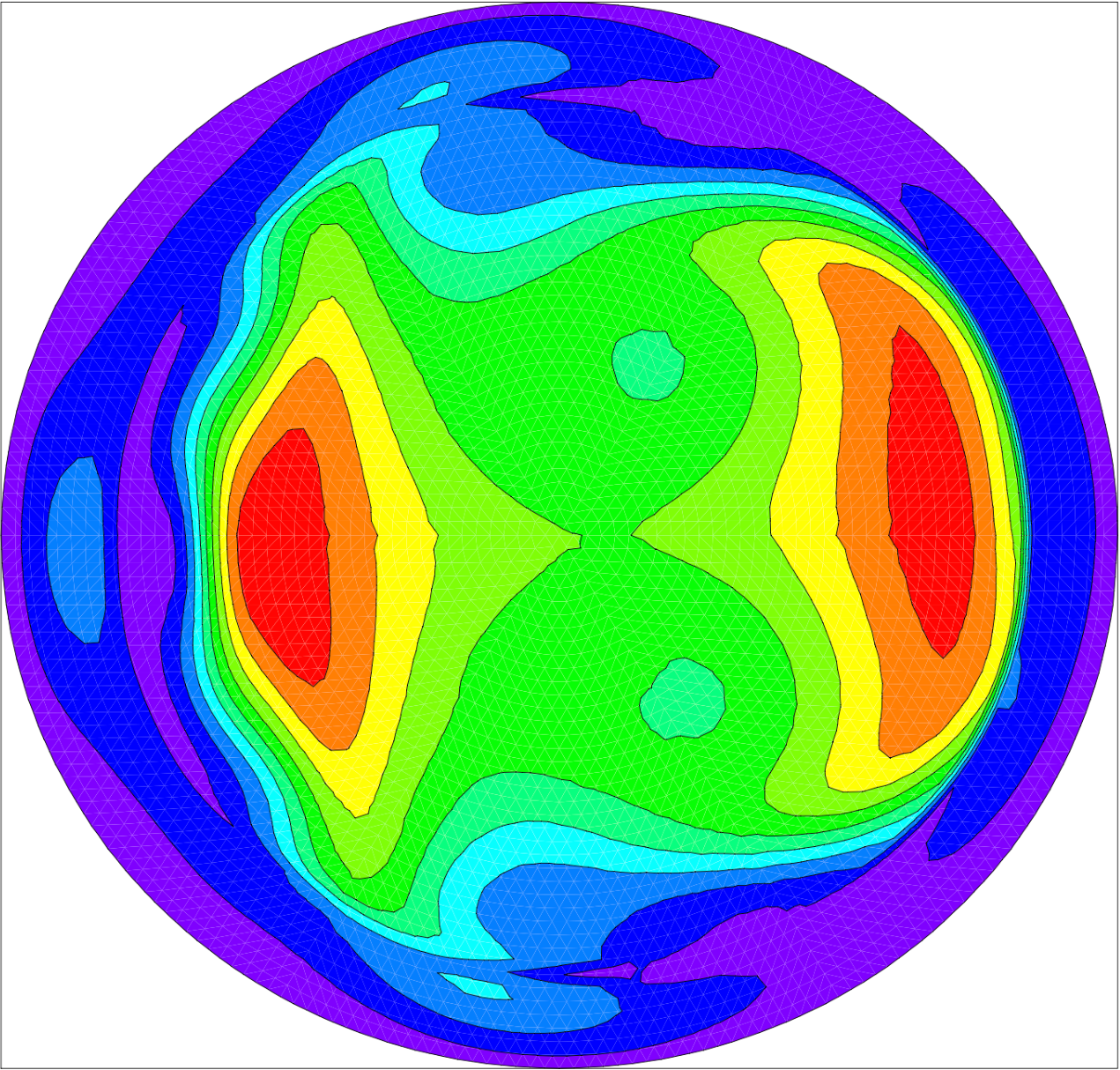}(a)
\includegraphics[width=4.3cm]{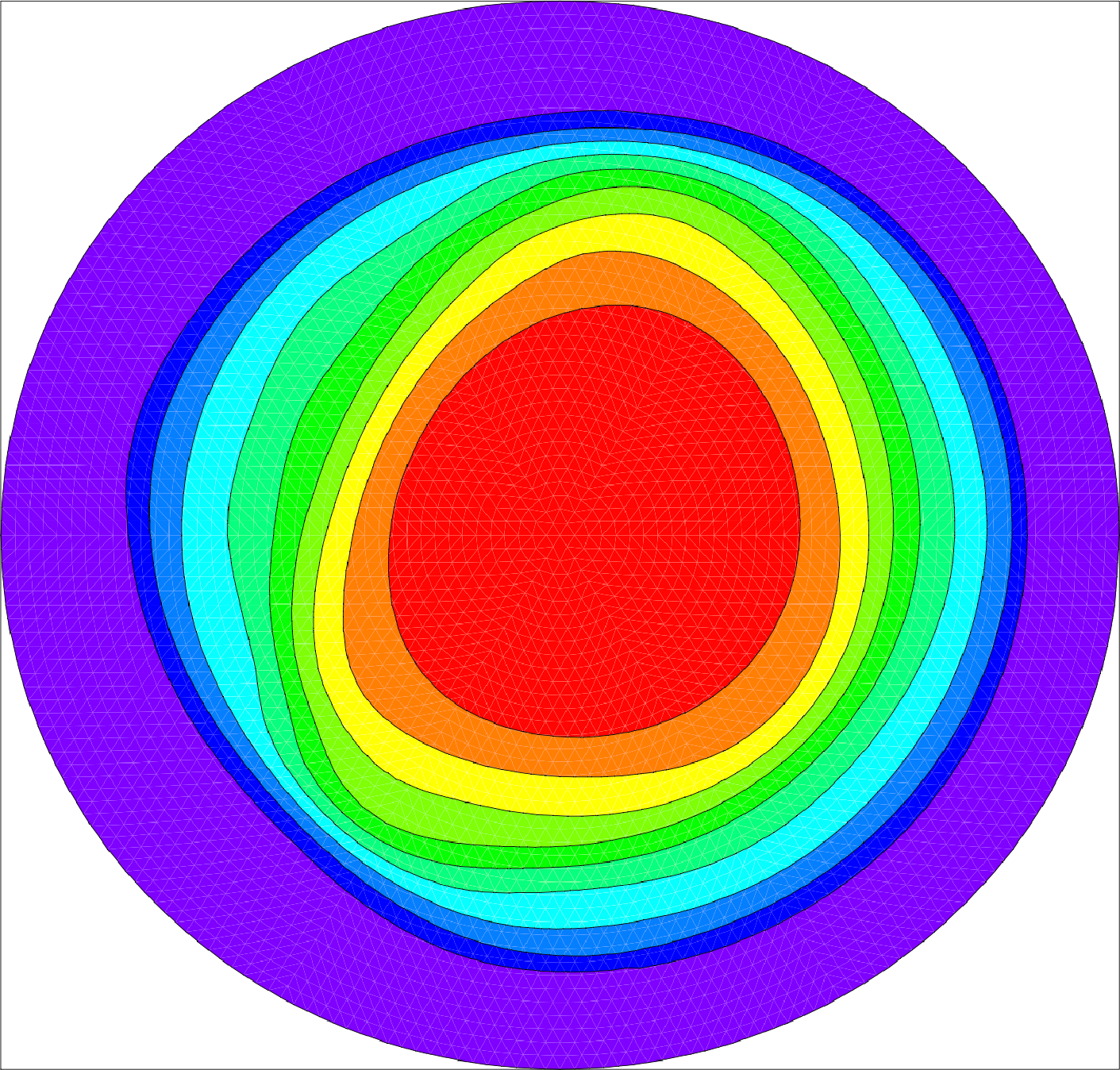}(b)
\end{center}
\caption{\it 
\Rdd{Simulations of cases in \rfig{qa3} with $q_a = 3$ and }
 (a) resistive wall showing pressure  $p$ contours with large $(2,1)$ island structure,
(b) ideal wall $p$ contours with small $(2,1)$ amplitude. 
}
\label{fig:qa2}
\end{figure}

Contours of pressure $p$ are shown in \rfig{qa2}, comparing two cases with $q_a = 3.$
In \rfig{qa2}(a), the wall is resistive, which in \rfig{qa2}(b),
the wall is ideal. The perturbations of $p$ are much larger in the resistive wall case.
Major disruptions only occur with a resistive wall.

\Grn{The major disruption onset limit $q_a < 3.4$ in \rfig{qa3}(b) depends on the amount of 
current contraction. \rfig{qa3}(a) shows that the $q = 2$ rational surface moves inward
as $q_a$ increases.  
If the $q = 2$ rational surface is too far from the wall, 
RWTMs are stable \cite{model}. In that case, major
disruptions do not occur, a result noted in a DIII-D database \cite{d3d22,sweeney2017}.
%In the database, disruptivity is low if the $q = 2$ rational surface is less then
%$0.63$ of the wall radius, consistent with the results in \rfig{qa3}.
%Here the wall is at $x = \pm 0.48.$
The limiting value $q_a < 3.4$ depends on details of the equilibrium, which will be
investigated elsewhere.}

These results demonstrate that  
RWTMs can cause a TQ under typical tokamak conditions for disruptions.

\end{section}

\begin{section}{Feedback} \label{sec:feedback}

It was shown in the previous sections than with an ideal wall, only 
minor disruptions occurred. This suggests that active feedback could
make the wall effectively ideal and suppress major disruptions, even for \Grn{$q_a > 2.$}

Feedback experiments on DIII-D \cite{hanson}  and RFX - mod \cite{piovesan} 
showed stabilization of what was
thought to be a RWM. The same approach ought to work for  RWTMs.

\begin{figure}[h]
\vspace{.5cm}
\begin{center}
\includegraphics[width=7.25cm] {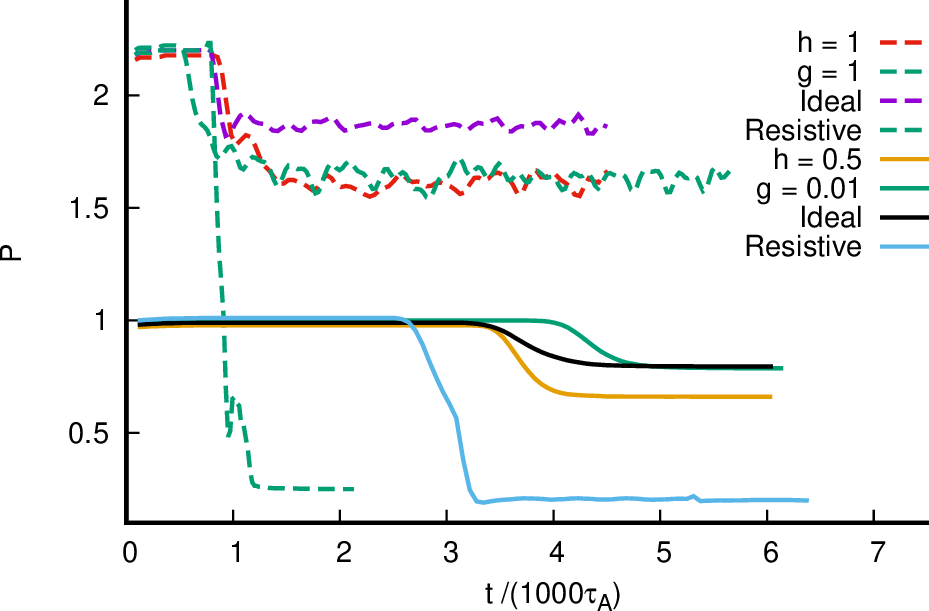}
\end{center}
% \vspace{-.5cm}
\caption{\it
Cases with $q_a = 2$ (dashed  curves) and $q_a = 3$ (solid curves)  
from  \rfig{qa3}.  Curves are plotted for ideal and resistive wall
\Grn{without feedback from \rfig{qa3}(b)} 
 and with feedback $h = 1, 0.5$ with $g = 0$;
and 
 $ g / S_{wall} = 1, 0.01,$ with $h = 0.$ These values prevent major disruptions. 
\Grn{The TQ time of major disruptions slows down as $q_a$ in increases. This
suggests that higher $q_a$ is easier to stabilize than $q_a = 2.$}
 }
\label{fig:fdbk}
\end{figure}
%\vspace{.5cm}

There have been extensive theoretical studies of feedback stabilization \cite{
gimblett, bondeson, liu,brennan}.
To model feedback, consider the magnetic diffusion equation at a thin
resistive wall \cite{jet21,d3d22,finn95}
\be %\gamma \psi 
\frac{\pd \psi_w}{\pd t} 
= \frac{\eta_w}{\delta_w} ( \psi'_{vac} - \psi_p' ) \label{eq:thin0} \ee
where $\psi_w$ % \propto \exp(\gamma t + im\theta + in \phi) $ 
is the magnetic potential at the wall,
$\psi_p'$ is its radial derivative on the plasma side of the wall,
$\eta_w,\delta_w$ are the wall resistivity and thickness,
and $\psi'_{vac}$ is the radial derivative of $\psi_w$ on the vacuum
side of the wall. 
The vacuum field is taken of the form
\be \psi_{vac} = \psi_w \left(\frac{r_w}{r}\right)^m
  + \psi_f \left[ \left(\frac{r_w}{r}\right)^m - \left(\frac{r}{r_w}\right)^m \right] 
\label{eq:psivac} \ee
where $ \psi_f = g D\psi_w/2 - hr_w F\psi_p'/(2m)$ is the feedback signal, $g$ is the normal gain,
%where $\psi_f = (g \psi - h (r_w / m) \psi')/2$ is the feedback signal, $g$ is the normal gain,
$h$ is the transverse gain,
$D(\theta,\psi_w), F(\theta,\psi_w)$ are screening functions of poloidal and
toroidal angle of the wall, \Grn{modeling  the location of the sensors,}
and $r_w$ is the wall radius.
For now, take $D = F =1.$
They \Grn{could}  be taken non zero in future numerical studies,
\Grn{might affect detailed predictions of the modeling.}
To obtain $g$, saddle coils which sense $b_n \propto \psi_w$, are required, 
which is fed back into $\psi_f.$
The measurement can be outside the wall, by continuity of $b_n.$
To obtain $h$, probes which sense transverse perturbed magnetic field $b_l \propto \psi_p'$ 
inside the wall are required,
and fed back into $\psi_f.$ 
Saddle coil sensors were used in RFX - mod \cite{zanca}, filtering the aliasing error.
Probes were used in DIII-D \cite{hanson}.

Then \req{thin0},\req{psivac}  can be expressed
\be \frac{\pd \psi_w}{\pd t} = -\frac{m }{\tau_{wall}} [  
(1 - h) \psi_p' +  (1+ g) \psi_w / r_w ] .
%- \Omega_w \frac{\pd \psi }{\pd \phi} 
\label{eq:thin1} \ee
Note that $\psi_w$ tends to an ideal wall, $\psi_w = 0,$  if
$ h = 1. \label{eq:hzero} $
%\be h = 1. \label{eq:hzero} \ee

The linear  dispersion relation \cite{d3d22} \req{disp1}  becomes
\be c_1 {S}^{3/4} S_{wall}^{-5/4} \hat{\gamma}^{5/4}  = \Delta_{i} + \frac{(1 -h)\Delta_{x}}
{\hat{\gamma}/g_s  +  1 + g } \label{eq:disp2}  \ee
where $\Delta_x = \Delta_n - \Delta_i.$ When $h = 1,$
the dispersion relation is that of an ideal wall. The $g$ term has to be much
larger to have a similar effect.

Nonlinear M3D simulations were carried out using \req{thin1} including $g,h$ applied
only for $n = 1$ toroidal harmonics. The equilibria of 
\rfig{qa3} were used.
Shown in \rfig{fdbk}  are the total pressure $P$ as a
function of time for ideal wall, resistive wall, and $g,h$ stabilization for
the cases $q_a = 2$ (dashed curves)  and $q_a = 3$ (solid curves)  of  \rfig{qa3}.
%The notation fdbk $k$ means $h = k.$
\Grn{Cases without feedback are reproduced from \rfig{qa3}(b) for comparison.}
For $q_a = 2,$ a major disruption is avoided  applying  $h=1,g=0$ feedback, while   
a similar result is obtained for $q_a = 3$ with $h = 0.5, g = 0.$
Major disruptions are avoided with $g / S_{wall} = 1, 0.01$, $h = 0.$
\Grn{The TQ time of major disruptions slows down as $q_a$ in increases.} This
suggests that feedback is more effective for larger $q_a$ than for
$q_a = 2.$  It appears more effective using $h$ than $g.$ 
Using $h = 1$ is not as effective as an ideal wall.
When  feedback experiments \cite{hanson,piovesan} were performed, it was not known that
RWTMs can cause disruptions.  It would be desirable to repeat the experiments
at higher $q_a$ to try to prevent disruptions.
Further results will be reported elsewhere.

Complex feedback
%\cite{gimblett, bondeson, betti} %. The result \req{gi} is equivalent to rotating the
will be studied elsewhere.
%plasma just inside the wall.

\end{section}
  \begin{section}{Conclusion} \label{sec:conclusion}

This article discussed disruptions caused by RWTMs. 
Linearly, RWTMs have growth times scaling as $S_{wall}^{\alpha}.$ 
For moderate $S_{wall},$ such as in JET and DIII-D, $\alpha < 1.$
For large $S_{wall}$, as in MST, $\alpha = 1,$ which could also be the case in ITER.
The thermal quench time in disruptions caused by RWTMs is proportional
to the linear growth time, which can be orders of magnitude longer for
large $S_{wall}$ than for moderate $S_{wall}.$ The large $S_{wall}$ growth time
has the same scaling as a RWM, and in simulations of MST  
the RWTM and RWM connect smoothly, as the $q = 2$
rational surface moves out of the plasma.

Simulations of a DIII-D disruption \cite{d3d22} were consistent with a RWTM. The mode growth rate,
thermal quench time, and model amplitude agreed with experiment.
Simulations of MST \cite{mst23} found a RWM  scaling of growth time and 
the nonlinear thermal quench time.

RWTM unstable equilibria have characteristic $q$ and current profiles,
in which the $q = 2$ rational surface is near the plasma edge, and the edge
current density is small. It was noted that common precursors to disruptions
can cause low edge temperature, which in turn causes low edge current density. 
The precursors could be overlapping tearing modes
or impurity radiation. The latter could cause density limit disruptions
by RWTMs. 
It was found in simulations of a sequence of low edge current equilibria, that a resistive
wall is required for a major disruption, indicating RWTMs. With an ideal
wall, there can be minor disruptions. It was found that there is an edge $q_a$
limit for major disruptions, $q_a \approx 3.$ This is typical of empirical
tokamak operating limits.

The ideal wall stability  suggests that active feedback
could prevent major disruptions. Simulations indicate that this is the case.
Feedback from probes and saddle coils was modeled, and it appears that probe
feedback is more effective. It also was found that it is harder to stabilize 
$q_a = 2$ than  $q_a = 3$ equilibria.  This is encouraging for feedback  experiments
to suppress major disruptions.

The RWTMs and RWMs considered here have low $\beta_N.$ The more well studied high
$\beta_N$ RWMs will be likely to have a high $\beta_N$ RWTM counterpart, which
will be studied in the future.

{\bf Acknowledgement} This work was supported by  U.S. D.O.E. grants DE-SC0020127, 
 DE-SC0020245 and DE-SC0019003.
\end{section}
%  \begin{section}{}

%\end{section}
\end{document}